 \documentclass[preprint, nopacs,titlepage, amsmath,superscriptaddress,longbibliography]{revtex4-2}
\usepackage{dcolumn}
\usepackage{bm}
\usepackage{graphicx}
\usepackage{color}
\usepackage{subfigure}
\usepackage{hyperref}
\usepackage{latexsym}
\usepackage{amsthm}
\usepackage{amssymb}
\usepackage{bm}
\usepackage{amsmath}

\usepackage{color}
\usepackage{ragged2e} 
\usepackage{lineno}
\usepackage{flushend}
\DeclareGraphicsExtensions{.jpg,.pdf, .mps, .png, .eps, .ps, .EPS,.gif}

\begin{document}
\def\be{\begin{equation}}
\def\ee{\end{equation}}

\def\bc{\begin{center}}
\def\ec{\end{center}}
\def\bea{\begin{eqnarray}}
\def\eea{\end{eqnarray}}
\newcommand{\avg}[1]{\langle{#1}\rangle}
\newcommand{\Avg}[1]{\left\langle{#1}\right\rangle}

\def\ie{\textit{i.e.}}
\def\etal{\textit{et al.}}
\def\m{\vec{m}}
\def\G{\mathcal{G}}

\newcommand{\gin}[1]{{\bf\color{blue}#1}}

\title{Topological Dirac cluster synchronization on directed hypergraphs}

\author{{Yupeng} {Guo}}
\affiliation{ School of Artificial Intelligence and Automation, Institute of Medical Equipment Science and Engineering, and State Key Laboratory of Digital Manufacturing Equipments and Technology, {Huazhong University of Science and Technology, {Wuhan}, {China}}}
 \author{{Ahmed A. A.} {Zaid}}
 \affiliation{School of Mathematical Sciences, Queen Mary University of London, London, E1 4NS, United Kingdom}
 \author{{Xueming} {Liu}\footnote{Email:xm\_liu@hust.edu.cn}}
 \affiliation{ School of Artificial Intelligence and Automation, Institute of Medical Equipment Science and Engineering, and State Key Laboratory of Digital Manufacturing Equipments and Technology, {Huazhong University of Science and Technology, {Wuhan}, {China}}}

\author{Ginestra Bianconi\footnote{Email:ginestra.bianconi@gmail.com}}
\affiliation{School of Mathematical Sciences, Queen Mary University of London, London E1 4NS, United Kingdom}

\begin{abstract}
Topological higher-order synchronization reveals collective phenomena demonstrating how topology shapes dynamics, yet providing a general theoretical framework for designing and controlling dynamical states on higher-order networks remains a challenge of the field. Here we propose a dynamical systems framework for  topological  cluster synchronization on directed hypergraphs that can be used to design synchronization patterns on nodes and hyperedges of directed hypergraphs. By encoding oriented hyperedges into the hypergraph topological  Dirac Hamiltonian, we obtain a spectrum whose isolated eigenstates correspond to distinct topological synchronization cluster states defined jointly on nodes and hyperedges. By selecting any isolated  eigenstate, the system can be driven toward the  associated dynamical state reflecting a specific partition of the hypergraph without modifying the underlying hypergraph structure. We numerically demonstrate the ability to design different topological cluster synchronization states on directed-hypergraph block models and empirical systems—including higher-order contact networks and the ABIDE functional brain network. Our results  establish a general and interpretable route for controlling collective dynamics in directed higher-order systems.
\end{abstract}

\maketitle

\section{Introduction}
 Higher-order networks~\cite{battiston2020networks,bianconi2021higher,bick2023higher} reveal a strong interplay  between topology and dynamics~\cite{millan2025topology,abiad2026hypergraphs,majhi2022dynamics,battiston2021physics} that significantly changes our understanding of network dynamics. From this growing body of work, it is emerging that higher-order dynamics unfolds not only on nodes, but also on edges and higher-dimensional structures of simplicial complexes and hypergraphs~\cite{millan2025topology}. Already at the node-based level, higher-order interactions have revealed important effects in spreading processes~\cite{iacopini2019simplicial,de2020social}, synchronization~\cite{skardal2020higher, millan2020explosive, wang2025higher, kundu2022higher}, random walks~\cite{schaub2020random,carletti2021random}, and percolation phenomena~\cite{sun2021higher,sun2023dynamic}. However, these node-based dynamical processes neglect higher-order topological dynamics  that appear in a large variety of scenarios, including  biological transport systems, brain connectivity~\cite{faskowitz2022edges,santoro2023higher,barbarossa2020topological}, and ecological flows.  Understanding these systems requires  embracing a topological description of the higher-order dynamics, in this way greatly complementing the power of topology in extracting information from higher-order structures~\cite{wei2025persistent,wang2023persistent}. 
Surprisingly, higher-order topological dynamics leads to collective phenomena such as topological synchronization~\cite{millan2020explosive,ghorbanchian2021higher,carletti2025global,bavcic2025phase,arnaudon2022connecting} and topological pattern formation~\cite{muolo2024turing} that have no equivalence when only node-based dynamics is considered.

The topological Dirac operator~\cite{bianconi2021topological} has recently emerged as a powerful tool for analyzing such higher-order topological interactions~\cite{giambagli2022diffusion,muolo2024three}. By coupling signals across adjacent dimensions of a simplicial complex or a hypergraph, the topological Dirac operator  is not only relevant to topological data analysis~\cite{wee2023persistent,suwayyid2024persistent}, but plays also a key role in determining  higher-order topological dynamics. Indeed, it provides   a principled framework for studying topological signal processing~\cite{calmon2023dirac,wang2025dirac}, machine-learning representations~\cite{battiloro2024generalized,nauck2024dirac,wee2023persistent}, synchronization dynamics~\cite{calmon2022dirac,calmon2023local,nurisso2024unified,carletti2025global,muolo2026synchronization}, and pattern formation~\cite{giambagli2022diffusion}. A notable development demonstrated that the eigenstates of the Dirac operator can be directly used to achieve topological cluster synchronization states on undirected graphs, through the Dirac-Equation Synchronization Dynamics (DESD) framework~\cite{zaid2025designing}. This result, derived for simple networks, establishes   a theoretical framework for designing dynamical patterns without modifying the underlying network structure and opens broad perspectives for controllability of higher-order networks.

However, higher-order topological dynamics has been, so far, only investigated in simple networks and simplicial complexes with undirected interactions for which algebraic topology is naturally defined.
Capturing higher-order  dynamics on directed higher-order networks is still in its infancy~\cite{gallo2022synchronization,gong2024higher}, despite the fact that many natural and engineered systems are governed by directed 
and higher-order communication patterns. Directed hypergraphs naturally capture these behaviors and therefore formulating higher-order topological dynamics on such structures constitutes an important gap in our understanding of higher-order topological dynamics.

In this work, our goal is to fill this gap by adopting the recent \cite{jost2019hypergraph} algebraic topology formulation of directed hypergraphs encoded in a single hypergraph boundary operator. While this algebraic topology framework has been used so far only for  defining node-based hypergraph dynamics, here we use this framework to define higher-order topological dynamics on hypergraphs. Specifically, we formulate the Dirac-Equation Synchronization Dynamics (DESD) on \emph{directed hypergraphs}, enabling the design of multiple topological cluster synchronization states on both nodes and hyperedges. Specifically, topological cluster synchronization implies that clusters of nodes and clusters of hyperedges oscillate with different frequencies thus inducing an effective partition of the hypergraph.  

By generalizing the hypergraph boundary matrix to obtain a degree-balanced operator, we construct the hypergraph Dirac operator whose spectral properties allow us to select specific topological cluster synchronization patterns. Indeed, by simply choosing a target isolated eigenstate of the Dirac Hamiltonian, the system self-organizes into the corresponding cluster state without modifying the underlying hypergraph. This generalization yields a family of synchronization patterns determined purely by the higher-order topology. We validate this mechanism through extensive simulations on directed-hypergraph stochastic block models and two empirical systems-the \textit{contact-high-school} network and the \textit{ABIDE} functional brain network~\cite{Mastrandrea-2015-contact,di2014autism,schaefer2018local,abraham2014machine}. In all cases, the DESD trajectories converge to the theoretically predicted clusters, demonstrating that spectral selection alone governs the accessible synchronization patterns in directed higher-order structures. Our results provide a general and interpretable route for engineering collective dynamics in directed hypergraphs, bridging algebraic topology, dynamical systems and network science.

\section{Results}
\subsection{Topological representation of directed hypergraphs}
\begin{figure*}[!ht]
	\centering
	\hspace{-0.5cm}
	\includegraphics[width=0.6\textwidth]{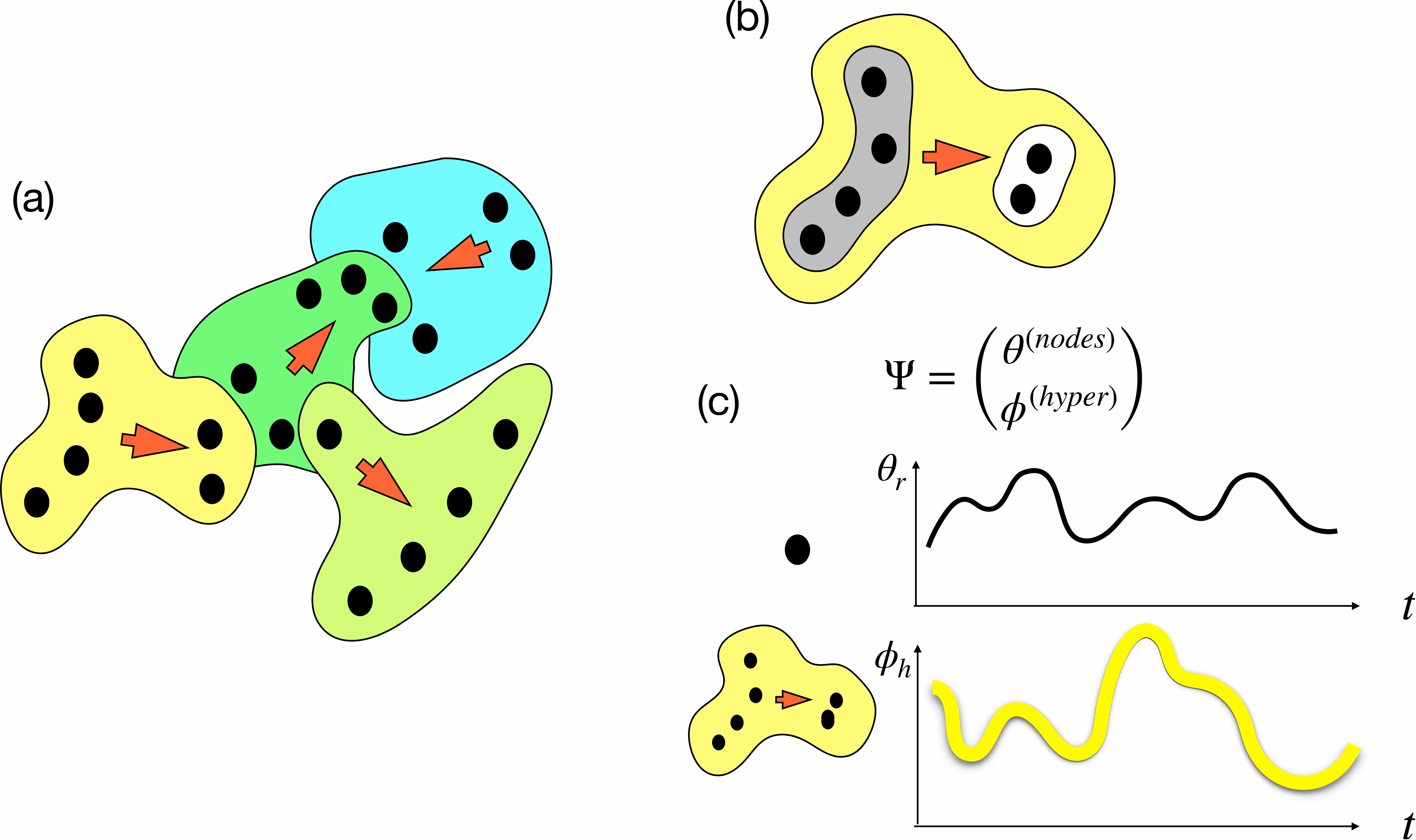}
	\vspace{-0.1cm}
	\caption{{ \textbf{An illustrative directed hypergraph.} 
(a) Schematic of a directed hypergraph with four directed hyperedges. (b) Each directed hyperedge is formed by a tail set (grey) and a head set (white). (c) The dynamics of directed hyperedges is captured by a topological spinor $\Psi$ comprising the dynamics associated with each node and the dynamics associated with each directed hyperedge of the hypergraph.}} 
	\label{HyperGraph}
\end{figure*}

Hypergraphs are flexible representations of higher-order networks comprising a set of $N_0$ nodes $V$ and a set of $N_1$ hyperedges, $\mathcal{E}$, mathematically capturing the higher-order interactions present in complex systems. Hypergraphs provide a very flexible structure that can  easily account for higher-order data, while simplicial complexes are the natural higher-order network representation that encodes higher-order network topology. Interestingly, however,  hypergraphs can also be amenable to topological treatment, either by treating them as weighted simplicial complexes without loss of information \cite{baccini2022weighted} or by directly formulating an algebraic topology of hypergraphs as recently proposed in Refs.\cite{jost2019hypergraph,mulas2020coupled}.   In this latter context, each hyperedge $h$ constitutes an ordered partition of nodes from a tail set, $T_h$, to a head set, $H_h$, thus defining a hyperedge orientation, as shown in Fig. \ref{HyperGraph}. 
 Similarly, each node $r$ is associated to a tail degree $k_r^{T}$ indicating the number of hyperedges  which include the node in their tail set, and to a head degree $k_r^{H}$ indicating the number of hyperedges  which include the node in their head set. In Ref. \cite{jost2019hypergraph} the topology of such oriented hypergraphs has been proposed and developed by mathematically establishing the properties of its associated boundary operator. By interpreting the hyperedge orientation with a hypergraph direction, this framework has been  applied to treat directed hypergraphs including most notably chemical reaction networks.  Here we build on these results and we propose to capture the topology of  oriented hypergraphs with the {\em normalized hypergraph boundary matrix} $B \in \mathbb{R}^{N_0\times N_1}$, with   elements  given by
\begin{equation}
	B_{rh} \;=\;
	\begin{cases}
		\;1/|H_h|, & r\in H_h,\\[2pt]
		-\;1/|T_h|, & r\in T_h,\\[2pt]
		0, & \text{otherwise}.
	\end{cases}
	\label{eq:B}
\end{equation}

This normalization makes the matrix degree-balanced, such that each column sums to zero ($\sum_{r \in V} B_{rh} = 0$). This implies that the constant vector ${1}_{N_0}\in \mathbb{R}^{N_0}$, defined on the nodes of the hypergraph, is harmonic. Thus, this choice allows us to  generalize the corresponding property of the harmonic eigenvector on simple networks,  i.e. ${ B}^{\top}{ 1}_{N_0}=0$. This does not exclude a priori the existence of other harmonic singular vectors ${ u}$ of ${ B}^{\top}$ but implies that any such vector needs to satisfy, for each hyperedge $h$, $\sum_{r\in H_h}u_{r}=\sum_{r\in T_h}u_r$. More broadly, this normalization keeps the contribution of each hyperedge at the same scale, regardless of its size, and avoids large hyperedges dominating the Dirac spectrum.

The framework extends naturally to weighted hypergraphs, where the weight $w_{rh}$ measures how much node $r$ contributes to hyperedge $h$,  which is conceptually close to other existing definitions of weights for hypergraphs \cite{ferraz2026assigning}. Specifically in this case, the boundary matrix $B$ would have elements as $ B_{rh}=w_{rh}/(\sum_{s\in H_h}w_{sh})$ if  $r\in H_h$ and $B_{rh}=w_{rh}/(\sum_{s\in T_h}w_{sh})$ if $r\in T_h$. This preserves the degree-balance property while allowing each node to contribute to any given hyperedge with different weights. The present work focuses on the unweighted scenario while the extension to the  weighted case can be considered for future investigations.

While the unnormalized hypergraph boundary operator and its associated Laplacian have been adopted to define node-based global synchronization of hypergraphs in Ref. \cite{mulas2020coupled}, here our focus would be to build on the topology of hypergraphs to formulate a Kuramoto-like dynamics to design dynamical patterns on synthetic as well as on real directed hypergraphs.

\subsection{Topological Dirac operator and spinor formulation}

\begin{figure*}[!ht]
	\centering
	\hspace{-0.5cm}
	\includegraphics[width=\textwidth]{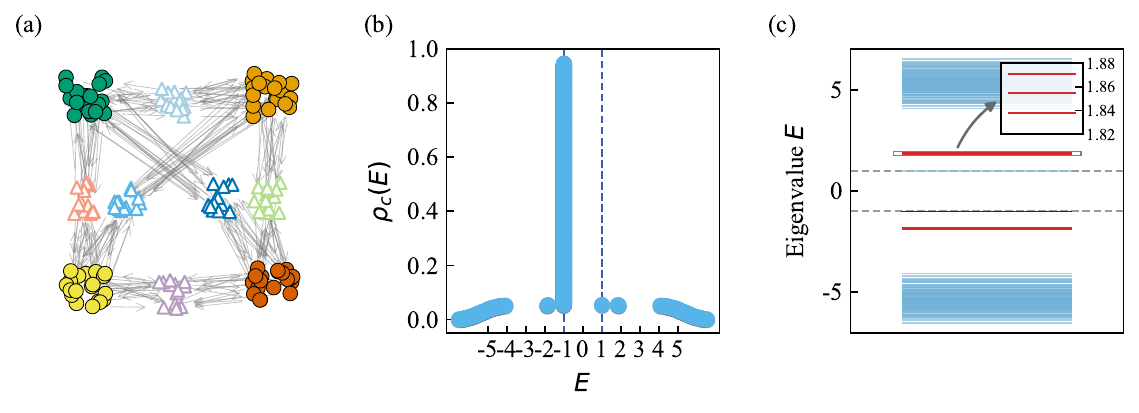}
	\vspace{-0.1cm}
	\caption{ \textbf{An illustrative directed hypergraph and its Dirac spectrum.}
		 (\textbf{a}) Schematic of the hypergraph, generated from a HSBM with a pronounced modular structure. The hypergraph consists of four blocks with $20$ nodes each (for a total of $V=80$ nodes). All hyperedges are uniform with a cardinality of $k_{\text{intra}}=k_{\text{inter}}=3$. The connectivity is governed by the expected intra-block average degree ($c_{\text{intra}}=50$) and inter-block average degree ($c_{\text{inter}}=1$), resulting in dense blocks (modules) with sparse connections between them. Nodes are represented by differently coloured circles and the oriented hyperedges by differently coloured triangles.  {(\textbf{b}) Cumulative density of states $\rho_c(E)$ indicating for positive value of $E$ the fraction of eigenstates with energy greater than $E$ and for negative value of $E$ the fraction of eigenstates with energy less than $E$. (\textbf{c}) Discrete energy-level diagram of the spectrum of the  Dirac Hamiltonian $\mathcal{H} = D + m\gamma$ with $m = 1$, for the hypergraph shown in panel~(\textbf{a}). Each eigenvalue is displayed as a horizontal line at its corresponding energy. Regions of higher intensity correspond to multiple eigenvalues with similar energies. The spectrum features a prominent gap in the interval $(-m, m)$ induced by the mass term. Isolated eigenvalues (red) are clearly separated from the bulk bands (blue).}
}
	\label{Fig.2}
\end{figure*}

Our goal is to design higher-order dynamical states, which go beyond node-based dynamics adopting the hypergraph topology of directed networks. 
To this end, we consider a topological description of the hypergraph dynamics which involves coupled topological signals $\theta\in \mathbb{R}^{N_0}$ and $\phi\in \mathbb{R}^{N_1}$ defined respectively on nodes  and hyperedges ~\cite{bianconi2021higher,millan2020explosive,ghorbanchian2021higher,zaid2025designing}. Thus, the dynamical state of a hypergraph is encoded in  a single  vector, the \emph{topological spinor}: $\Psi=(\theta,\phi)^\top\in \mathbb{R}^{N_0+N_1}$, where the coupling is mediated by the topological Dirac operator, $D$ \cite{bianconi2021topological}. Here, this operator is constructed directly from the hypergraph boundary matrix (Eq.~\ref{eq:B}) as:
\begin{equation}
	D \;=\;
	\begin{pmatrix}
		0 & B\\
		B^{\top} & 0
	\end{pmatrix}.
	\label{eq:D}
\end{equation}
The cornerstone of our control scheme is the \emph{Dirac Hamiltonian}, $\mathcal{H}$, which introduces a mass term $m$ and a chiral symmetry operator $\gamma$ to the system~\cite{ bianconi2021topological,zaid2025designing,wang2025dirac}:
\begin{equation}
	\mathcal{H} \;=\; D + m\,\gamma, \qquad \text{where} \qquad \gamma \;=\; \begin{pmatrix}{I}_{N_0}&0\\0&-{I}_{N_1}\end{pmatrix},
	\label{eq:H_final}
\end{equation}
where $I_N$ indicates the $N\times N$-identity matrix.
A key consequence of this Hamiltonian is its relativistic-like dispersion relation, $E^2 = \lambda^2 + m^2$, where $\lambda$ is a singular value of $B$~\cite{bianconi2021topological}.  This opens a spectral gap of width $2m$, i.e., $|E|\ge m$. For each nonzero $\lambda$, the spectrum contains a pair of eigenvalues $E=\pm\sqrt{\lambda^2+m^2}$, so all non-harmonic levels appear in symmetric pairs with $|E|>m$.
The corresponding eigenstates of $\mathcal{H}$ are the eigenstates of the  Topological Dirac equation. Specifically, denoting by ${ u}_{\lambda}$ and by ${ v}_\lambda$ the left and right singular vectors of ${B}$ respectively, the eigenstates associated with energies $|E|>m$  are given by \cite{bianconi2021topological,wang2025dirac,zaid2025designing}
\bea
{\Psi}^{(E)}&=&\mathcal{C}^+\left(\begin{array}{c}
        { u}_\lambda \\
        \frac{ \lambda}{|E|+m} { v}_\lambda
    \end{array}\right),\quad\mbox{for}\  E>m,\quad
 {\Psi}^{(E)}=\mathcal{C}^- \left(\begin{array}{c}
    \frac{ \lambda}{|E|+m} { u}_\lambda \\
       -{ v}_\lambda
    \end{array}\right),\quad\mbox{for}\  E<-m,
    \label{eq:Dirac_eigestates}
\eea
where $\mathcal{C}^{\pm}$ are normalization constants.
Thus, the mass modulates the relative normalization of node and hyperedge contributions. Specifically, for  large values of the mass $m$, the node signal contribution dominates for eigenstates associated to $E>m$ while hyperedge signal contribution dominates for eigenstates associated to $E<-m$. The  harmonic modes of energy $E=\pm m$ which correspond to the singular value $\lambda=0$ of the boundary operator, are instead exclusively localized on nodes, for $E=m$, or on hyperedges, for $E=-m$, i.e.
\bea
{\Psi}^{(E)}=\left(\begin{array}{c}
        { u}_0 \\
        { 0}
    \end{array}\right), \quad\mbox{for}\  E=m,
  \quad    {\Psi}^{(E)}=\left(\begin{array}{c}
        { 0} \\
        { v}_0 \\
    \end{array}\right),  \quad\mbox{for}\  E=-m,
    \nonumber
\eea
where ${u}_0$ and ${v}_0$ indicate the generic harmonic left and right singular vectors of the hypergraph boundary matrix $B$.
As it occurs also for networks, the degeneracies of the positive and negative harmonic eigenstates are in general different and are indicated with $\beta_0$ and $\beta_1$ respectively (see Fig.\ref{Fig.2}). In networks $\beta_0$ and $\beta_1$ are given by the Betti numbers of the networks: the number of connected components ($\beta_0$) and the number of independent cycles ($\beta_1$). In hypergraphs  $\beta_n$ indicates the {\it hypergraph $n$-Betti number}, a generalized notion of Betti numbers for the hypergraph topology. Since here we adopt a normalized hypergraph boundary matrix given by Eq.(\ref{eq:B}), these Betti numbers will depend non-trivially on the hypergraph topology under consideration.

 While the standard Kuramoto model \cite{rodrigues2016kuramoto} and many topological synchronization models \cite{millan2020explosive,ghorbanchian2021higher} focus on synchronization patterns aligned to the harmonic eigenstates, here our focus is on designing topological synchronization patterns aligned to non-harmonic eigenstates of energies $|E|>m$. In random higher-order networks, non-harmonic eigenstates typically exhibit a compact bulk spectrum. However, structural heterogeneities of the network like dense modules and a non-trivial community structure give rise to isolated eigenstates localized on the very clusters that generate them~\cite{hata2017localization, von2007tutorial}. For the Dirac spectrum, the eigenvectors corresponding to the isolated states of energy $|E|>m$ serve as ``spectral handles'' for targeted control. Our strategy leverages the synergy between these two features. First,  by targeting an isolated positive or negative eigenstate of the Dirac operator we can  jointly induce a functional and dynamical partition of the nodes and of the hyperedges of the hypergraph which is guaranteed to be stable.  Secondly,  by choosing different values of the mass $m$ we can modulate the relative contribution of the node and hyperedge signals.

\subsection{Dirac-Equation Synchronization Dynamics (DESD)}
The Topological Dirac operator and the spectral decomposition obtained by considering the Dirac Hamiltonian  provide a general algebraic topology treatment of directed hypergraphs that allows us to design topological Dirac cluster synchronization states.
These are topological dynamical states where different sets of nodes and different sets of hyperedges oscillate at the same frequency. Thus, topological Dirac cluster  synchronized states induce dynamical patterns that can dynamically partition the considered hypergraph in different ways. 
Beyond node-based dynamics, the framework of higher-order topological dynamics \cite{millan2025topology} will allow us to treat the dynamics  defined on both nodes and hyperedges. To this end we will  adopt a  higher-order topological Kuramoto-like approach \cite{millan2020explosive,ghorbanchian2021higher,calmon2022dirac}.
The evolution of the topological spinor $\Psi$ is governed by the Dirac-Equation Synchronization Dynamics (DESD), a Kuramoto-like model proposed in  Ref.~\cite{zaid2025designing} for pairwise networks and here adopted to describe the higher-order dynamics of hypergraphs:
\begin{equation}
	\dot\Psi \;=\; \Omega \;-\; \sigma\, (\mathcal{H}-\bar E I)\,\sin\!\big((\mathcal{H}-\bar E I)\Psi\big),
	\label{eq:desd}
\end{equation}
where $\Omega=(\omega_{nodes},\omega_{hyper})^{\top}\in \mathbb{R}^{N_0+N_1}$ is the vector of natural frequencies associated to the nodes ($\omega_{nodes}\in \mathbb{R}^{N_0}$) and the hyperedges ($\omega_{hyper}\in \mathbb{R}^{N_1}$), and $\sigma$ is the global coupling strength. 
This dynamics can be derived from the {\em free-energy} $\mathcal{F}$ given by 
\bea
\mathcal{F}=-{ 1}^{\top}_{N_1}\cos\!\big((\mathcal{H}-\bar E I)\Psi\big)
\eea
by implementing a dynamical gradient flow 
\bea
\dot\Psi= \Omega -\frac{\partial \mathcal{F}}{\partial \Psi}.
\eea
Let us decompose the dynamics into a component $\Psi_{\parallel}$ proportional to the eigenvector $\Psi^{(\bar{E})}$ and a component $\Psi_{\perp}$ orthogonal to it with $\Psi=\Psi_{\parallel}+\Psi_{\perp}$. Similarly, let us decompose the vector $\Omega$ of the intrinsic frequencies  as the sum of $\Omega_{\parallel}$, representing the component of $\Omega$ along the eigenstate of energy $\bar{E}$, and $\Omega_{\perp}$ representing $\Omega_{\perp}=\Omega-\Omega_{\parallel}$. In this way, it is easy to show that  DESD, dictated by Eq.(\ref{eq:desd}), decomposes into a free oscillation of $\Psi_{\parallel}$ with frequency $\Omega_{\parallel}$  and a coupled dynamics for $\Psi_{\perp}$:
\bea
\dot\Psi_{\parallel}&=&\Omega_{\parallel}\nonumber\\
\dot\Psi_{\perp}&=&\Omega_{\perp}-\sigma\, (\mathcal{H}-\bar E I)\,\sin\!\big((\mathcal{H}-\bar E I)\Psi_{\perp}\big)\label{eq:Psi_dec}
\eea
As $\sigma$ is raised,  $\Psi_{\perp}$ eventually reaches a stationary, frozen state and the dynamics is just encoded by the component $\Psi_{\parallel}$. Thus, in this limit, we find that the dynamics is  encoded in $\Psi_{\parallel}$, which evolves as
\bea
\Psi_{\parallel}(t)=\Psi_{\parallel}(0)+\Omega_{\parallel}t\simeq \Omega_{\bar{E}}t \Psi^{(\bar{E})},
\label{eq:synchr_dyn}
\eea
where here  $\Omega_{\bar{E}}=\Avg{\Omega,\Psi^{(\bar{E})}}$ and  $\Avg{a,b}$  indicates the scalar product between $a$ and $b$.
It follows that for large values of $\sigma$, as long as the DESD reaches a stable synchronized state, each topological oscillator $\mu$  (either associated with a node $\mu=r$ or with a hyperedge $\mu=h$) oscillates at a frequency proportional to the corresponding component $\Psi_{\mu}^{(\bar{E})}$ of the eigenstate associated with energy $\bar{E}$. Thus, by selecting different values $\bar{E}$ for the selected eigenstate, DESD provides  a powerful framework for designing  dynamical patterns on directed hypergraphs. The order parameters for the DESD dynamics are given by $R_0=|\sum_{r=1}^{N_0}e^{\textrm{i}\alpha_r}|/N_0,R_1=|\sum_{h=1}^{N_1}e^{\textrm{i}\beta_h}|/N_1$ with $(\alpha,\beta)^{\top}=(\mathcal{H}-\bar{E}I)\Psi$ and by $f=\mathcal{F}/(N_1)$. In particular, high values of $R_{0},R_1$ (i.e. $R_n\simeq 1$) and small values of $f$ (i.e. $f\simeq -1$) indicate that the DESD achieves a topological Dirac cluster synchronization state.
Having established that DESD admits a topological Dirac cluster synchronization state aligned along the eigenstate $\bar{E}$, the important theoretical question is to establish whether this state is stable. The linear stability analysis establishes that such a state is stable  if the eigenmode associated with the eigenstate $\bar{E}$ is isolated (see Methods).  %

\begin{figure*}[!ht]
	\centering
	\hspace{-0.5cm}
	\includegraphics[width=0.6\textwidth]{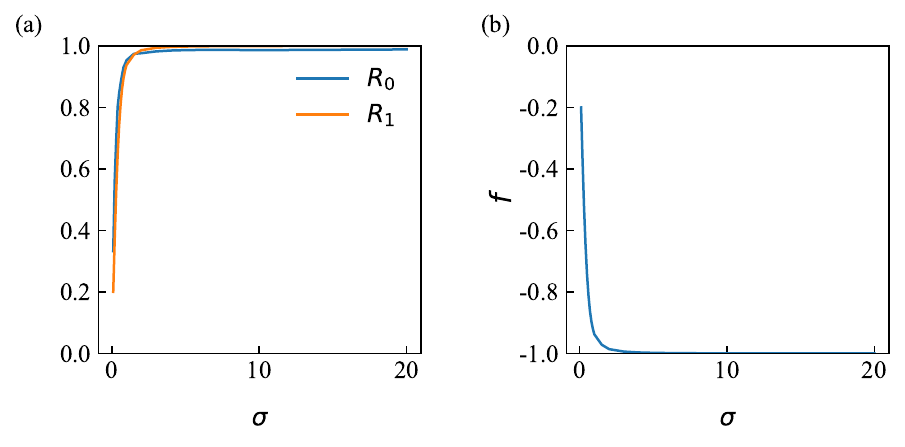}
	\vspace{-0.1cm}
    \caption{
    \textbf{Evidence of the synchronization of DESD.} 
 The DESD order parameters  $R_0$ and $R_1$ (panel (a)), and   $f$ (panel (b)), are plotted as functions of the coupling strength $\sigma$. The results are shown for a typical isolated eigenstate of the Dirac spectrum of the HSBM shown in Fig.~\ref{Fig.2}. Simulations are obtained along the forward transition by raising the coupling constant $\sigma$
 starting from initial conditions at $\sigma=0$. For
 each data point along the annealed dynamics, DESD is run up to a maximum time $T_{\textrm{max}}=30$ with $dt=5\times 10^{-5}$; values are averaged after equilibration.
 }
	\label{Orderpara}
\end{figure*}

\subsection{Spectral structure of the directed Hypergraph Stochastic Block Models (HSBM) }

\leavevmode\indent
We illustrate the link between mesoscopic topology and the  spectrum of the Dirac Hamiltonian using 
an illustrative directed hypergraph generated from a Hypergraph Stochastic Block Model (HSBM)~\cite{deng2023strong}. The hypergraph contains $V=80$ nodes arranged into four well-defined modules, with directed hyperedges preferentially formed within each community (Fig.~\ref{Fig.2}(a)).  {The spectrum of the corresponding  Dirac Hamiltonian $\mathcal{H} = D + m\gamma$, shown in Fig.~\ref{Fig.2}(b)-(c), exhibits a prominent gap in the interval $(-m, m)$ induced by the mass term. Several isolated eigenvalues are separated from the bulk spectrum; each corresponds to a Dirac eigenmode supporting a distinct topological cluster synchronization pattern.}

\begin{figure*}[!ht]
	\centering
	\hspace{-0.5cm}
	\includegraphics[width=0.7\textwidth]{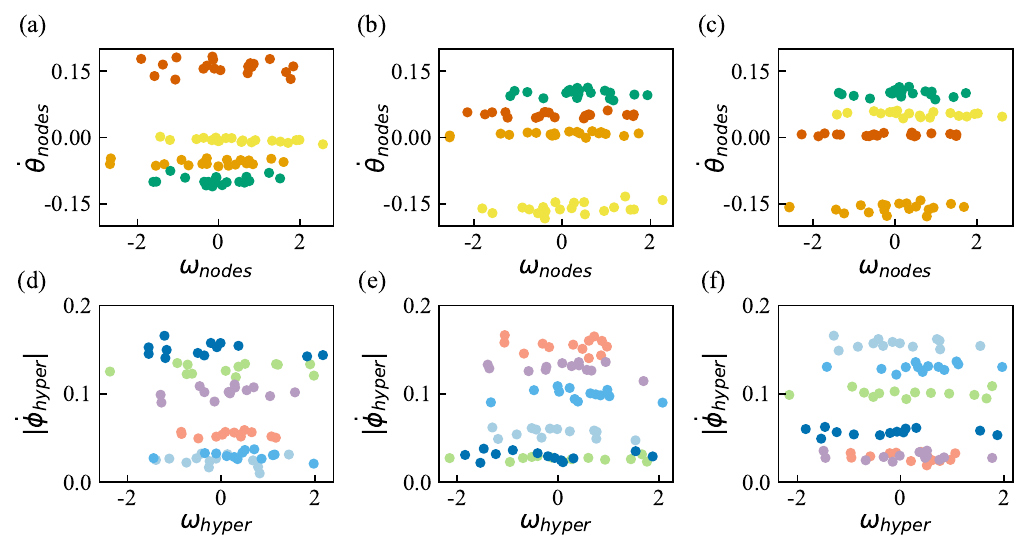}
	\vspace{-0.1cm}
    \caption{ {\textbf{Topological Dirac cluster synchronization states provide a flexible framework for the partition of directed hypergraphs.} For sufficiently high values of the coupling constant $\sigma$, the DESD dynamics achieves topological Dirac cluster synchronization where different clusters of  oscillators located on nodes oscillate at unison with frequencies  $\dot{\theta}$ (\textbf{a}-\textbf{c}), while different clusters of oscillators on  hyperedges oscillate at unison with frequencies $\dot{\phi}$ (\textbf{d}-\textbf{f}).  The hypergraph is the same as in Fig.~\ref{Fig.2} and colors correspond to the block structures indicated there.  By selecting different isolated eigenstates ${\bar{E}}$ of the Dirac Hamiltonian (corresponding to the three different columns) the topological Dirac cluster synchronization state changes, effectively inducing a different dynamical partition on nodes and hyperedges of the hypergraph. {The DESD dynamics is evolved up to $T_{\max}=30$ with mass $m=1$ and coupling strength $\sigma=1000$.} }
    }
	\label{Fig.3}
\end{figure*}

In Fig.~\ref{Orderpara}, we show evidence of the synchronization transition occurring for the  DESD dynamics designed to occur along an isolated eigenstate $\bar{E}$. As the coupling constant $\sigma$ is raised, the order parameters indicate the achievement of the topological cluster synchronization obtained when  $R_0, R_1 \to 1$ and $f \to -1$. Furthermore, the dynamics allows the flexible design of different cluster synchronized states, depending on the choice of the eigenstate $\bar{E}$ among the isolated eigenstates of the Dirac Hamiltonian (see Fig.~\ref{Fig.3}). Indeed, different eigenstates produce different topological cluster synchronization states that correspond to  different dynamical partitions of nodes and hyperedges of the hypergraph remaining compatible with the underlying block structure (shown in Fig.~\ref{Fig.2}).

\begin{figure}[!ht]
	\centering
	\hspace{-0.5cm}
	\includegraphics[width=0.5\textwidth]{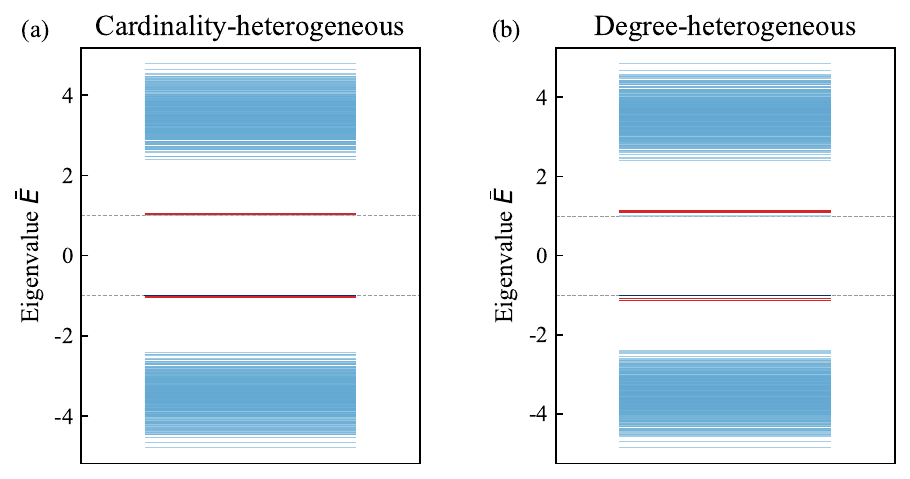}
	\vspace{-0.1cm}
    \caption{  { \textbf{Spectral comparison of two HSBM variants.} Energy spectrum of the Dirac Hamiltonian $\mathcal{H} = D + m\gamma$  (with $m = 1$) for \textbf{(a)} the cardinality-heterogeneous model ($c_{\text{intra}}=c_{\rm inter}=37$, $k_{\text{intra}}=4$, $k_{\text{inter}}=70$) and \textbf{(b)} the degree-heterogeneous model ($c_{\text{intra}}=37$, $c_{\text{inter}}=1$, $k_{\text{intra}}= k_{\rm inter}=4$). The cardinality-heterogeneous model exhibits spectral crowding, with isolated eigenvalues close to the bulk, whereas in the degree-heterogeneous model the isolated eigenvalues are well separated. }
}
	\label{Fig.3add}
\end{figure}

As  a synthetic test beds, we consider two classes of hypergraph stochastic block models (HSBMs), each introducing a distinct form of structural heterogeneity. We first consider the case of \emph{heterogeneous hyperedge cardinality} in which intra-block hyperedges have small cardinality $k_{\text{intra}}$, whereas inter-block hyperedges have large cardinality $k_{\text{inter}}$, with $k_{\text{intra}} \ll k_{\text{inter}}$.  
To isolate the effect of this modulation of the hyperedge cardinality, the expected node degree is kept constant across the hypergraph ($c_{\text{intra}}=c_{\text{inter}}$).  
This generates a characteristic architecture composed of dense intra-block connectivity via many small hyperedges and sparse inter-cluster connectivity mediated by fewer but much larger hyperedges. The second class of HSBMs introduces heterogeneity to average node degrees ($c_{\text{intra}} \gg c_{\text{inter}}$) while keeping the hyperedge cardinality fixed.  
This model represents a more conventional form of modularity~\cite{deng2023strong}.  
Here, isolated eigenvalues are typically well separated, and the DESD dynamics typically reaches the topological cluster synchronized state provided that the coupling constant $\sigma$ is large enough.

\begin{figure*}[!ht]
    \centering
    \hspace{-0.5cm}
    \includegraphics[width=0.6\textwidth]{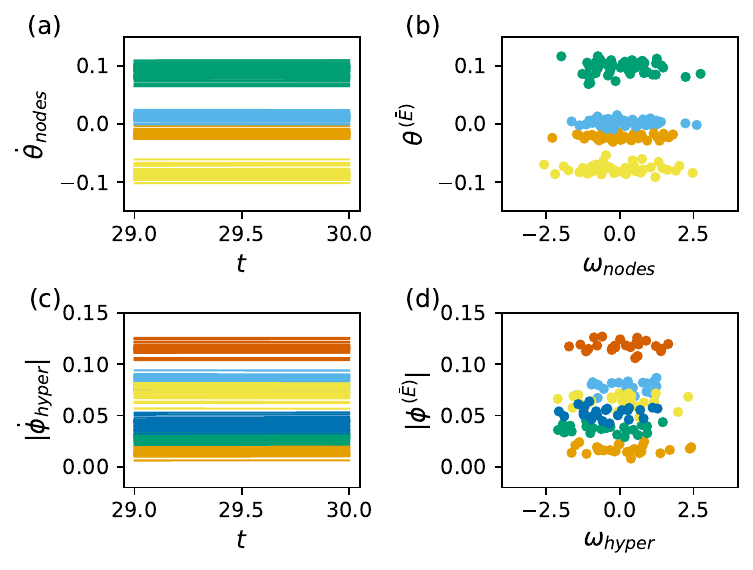}
    \vspace{-0.1cm}
    \caption{ \textbf{Spectral design of topological Dirac cluster synchronization states on a degree-heterogeneous HSBM.}  By selecting an eigenstate of energy $\bar{E}$ it is possible to design topological Dirac cluster synchronization states in which nodes (\textbf{a}) and hyperedges (\textbf{c}), asymptotically in time $t$ have frequencies $\dot{\theta}_{\textrm{nodes}}$ and $\dot{\phi}_{\textrm{hyper}}$  proportional to the elements of $\Phi^{(\bar{E})}=(\theta^{(\bar{E})},\phi^{(\bar{E})})^{\top}$ associated to nodes (\textbf{b}) and hyperedges (\textbf{d}).  The HSBM parameters are: $c_{\text{intra}}=50$, $c_{\text{inter}}=1$, $k=4$. Simulations are run from random initial conditions with parameters: $\sigma=270$, $m=1.0$, $T_{\max}=30$.}
	\label{Fig.5}
\end{figure*}

 {As shown in Fig.~\ref{Fig.3add}, the two models exhibit distinct spectral structures: in the cardinality-heterogeneous model, isolated eigenvalues lie close to the bulk, whereas in the degree-heterogeneous model they are well separated. When isolated eigenvalues are closely spaced, nearby non-target modes contribute to the DESD dynamics, resulting in slower convergence toward the target cluster state. As shown in Fig.~\ref{Fig.5}, simulations converge directly to a stable configuration that aligns with the theoretical prediction. {When isolated eigenvalues are well separated, as in the degree-heterogeneous HSBM, DESD converges directly to the target state. When they are closely spaced, as in the cardinality-heterogeneous HSBM, convergence is slower and requires raising the value of $\sigma$  by a factor $z$ or equivalently rescaling $\Omega_{\parallel}$ by $1/z$ and time by a factor $z$.}

\begin{figure*}[!ht]
    \centering
    \hspace{-0.5cm}
    \includegraphics[width=\textwidth]{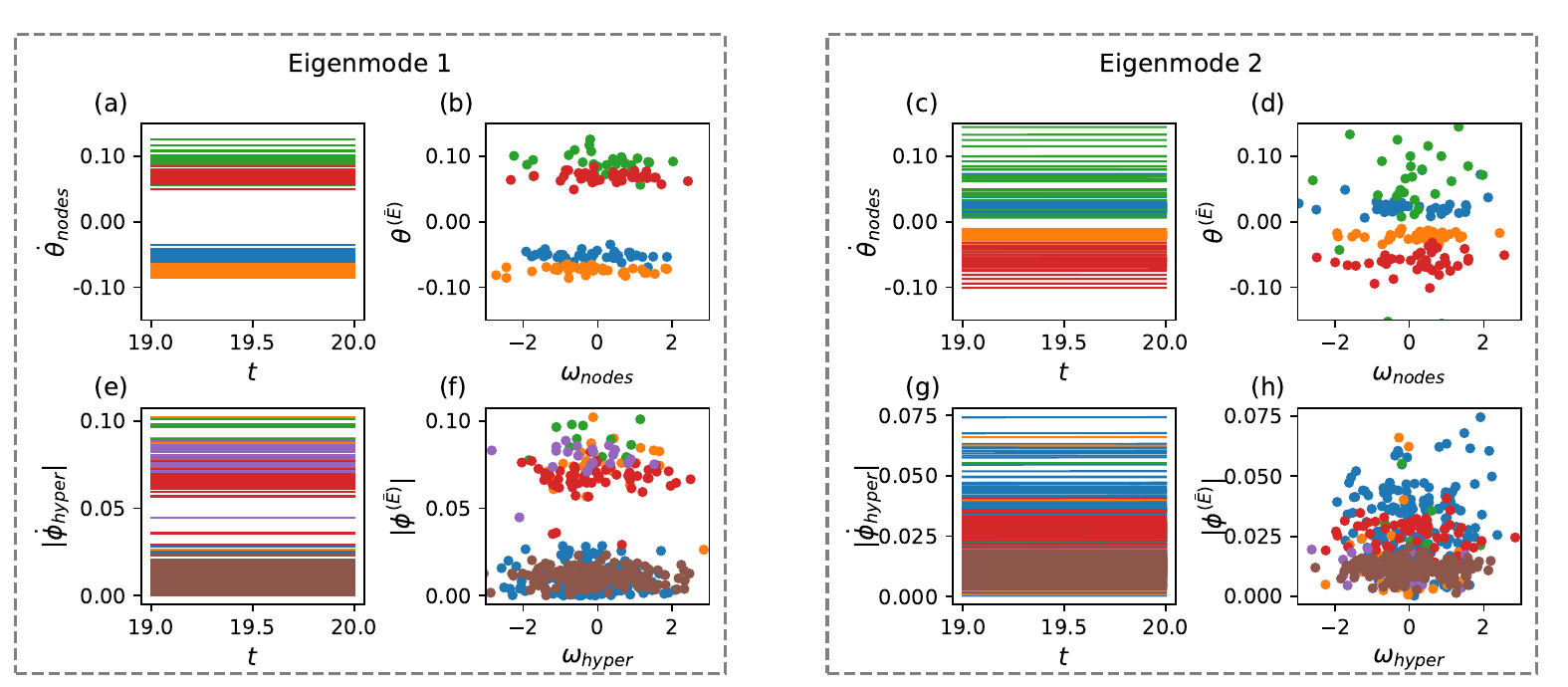}
    \vspace{-0.1cm}
    \caption{ \textbf{Evidence of at least two topological Dirac cluster synchronization states on the face-to-face contact hypergraph.} The left block (\textbf{a}, \textbf{b}, \textbf{e}, \textbf{f}) corresponds to Eigenmode 1, and the right block (\textbf{c}, \textbf{d}, \textbf{g}, \textbf{h}) to Eigenmode 2. For each eigenmode, the top row shows node frequencies $\dot{\theta}$ and the bottom row shows hyperedge frequencies $\dot{\phi}$ as a function of time $t$.  Panels (\textbf{b}, \textbf{d}) and (\textbf{f}, \textbf{h}) show the theoretical predictions from the target eigenstates $\theta^{(\bar{E})}$ and $|\phi^{(\bar{E})}|$ as a function of the intrinsic frequencies $\omega_{\textrm{nodes}}$ and $\omega_{\textrm{hyper}}$. The face-to-face contact network \cite{Mastrandrea-2015-contact}  has $N_0=150$ nodes, $N_1=3160$ hyperedges; groups: $\{\mathrm{PC}, \mathrm{PC}^*, \mathrm{MP}^*1, \mathrm{MP}^*2\}$. The simulation  parameters are: $m=1.0$, $\sigma=50$, $T_{\max}=20$.}
	\label{Fig.6}
\end{figure*}

\begin{figure*}[!ht]
    \centering
    \hspace{-0.5cm}
    \includegraphics[width=\textwidth]{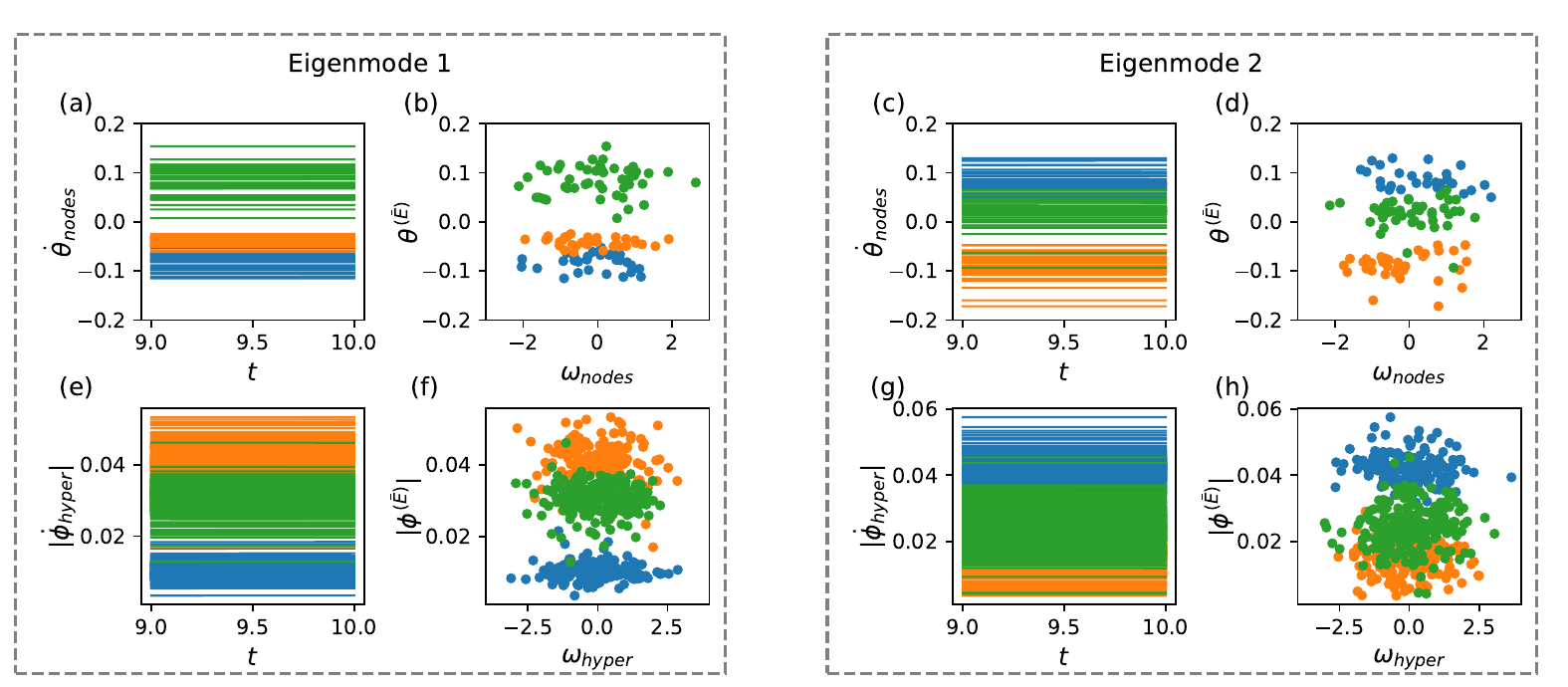}
    \vspace{-0.1cm}
    \caption{ \textbf{Evidence of at least two topological Dirac cluster synchronization states on the ABIDE  brain functional sub-hypergraph.}  The left block (\textbf{a}, \textbf{b}, \textbf{e}, \textbf{f}) corresponds to Eigenmode 1, and the right block (\textbf{c}, \textbf{d}, \textbf{g}, \textbf{h}) to Eigenmode 2. For each eigenmode, the top row shows node frequencies $\dot{\theta}$ and the bottom row shows hyperedge frequencies $\dot{\phi}$ as a function of time $t$.  Panels (\textbf{b}, \textbf{d}) and (\textbf{f}, \textbf{h}) show the theoretical predictions from the target eigenstates $\theta^{(\bar{E})}$ and $|\phi^{(\bar{E})}|$ as a function of the intrinsic frequencies $\omega_{\textrm{nodes}}$ and $\omega_{\textrm{hyper}}$.  The ABIDE  brain functional sub-hypergraph has $N_0=110$ nodes, $N_1=4235$ hyperedges; groups: Default, Visual, and Somatomotor networks. Hyperedges are defined by selecting the top 60\% of within-system and top 20\% of cross-system connections (correlation threshold $\approx 0.50$). The simulation parameters are: $m=1.0$, $\sigma=20.0$, $T_{\max}=10$.}
	\label{Fig.7}
\end{figure*}

\begin{figure*}[!ht]
    \centering
    \hspace{-0.5cm}
    \includegraphics[width=0.8\textwidth]{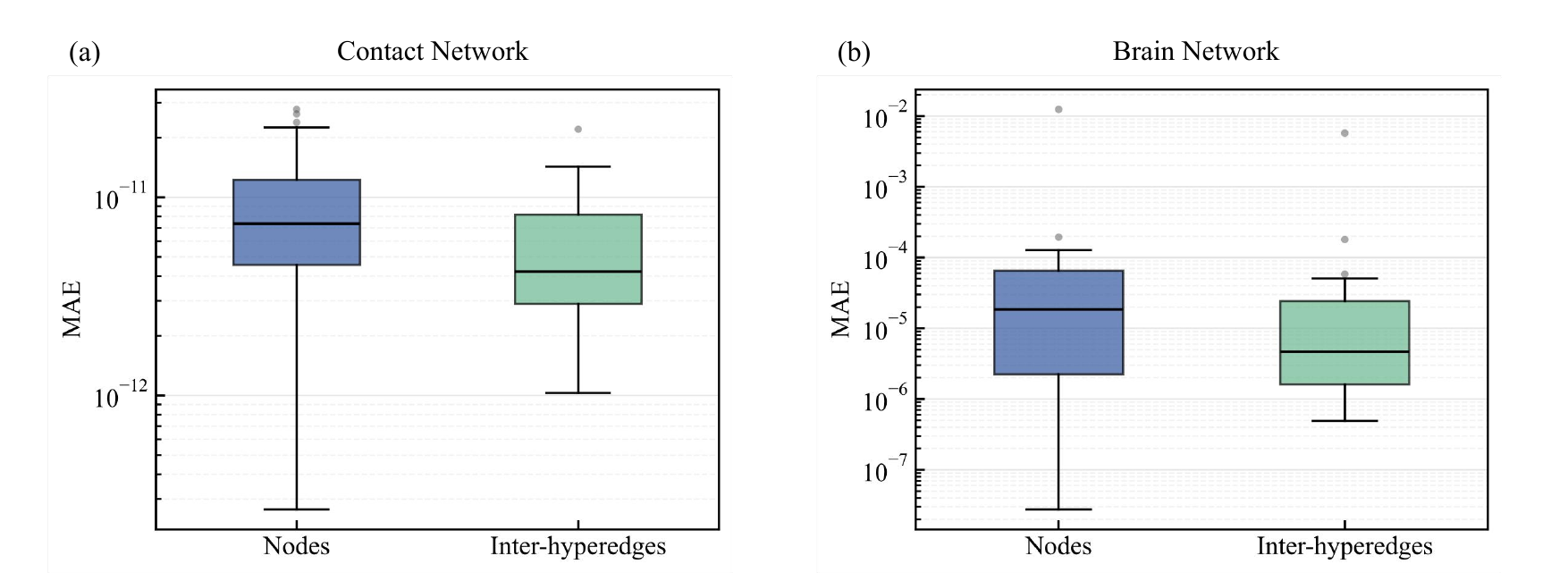}
    \vspace{-0.1cm}
    \caption{ 
     {\textbf{Robustness to random head-tail assignments.} The ability to achieve the topological Dirac cluster synchronization is independent of the random head-tail assignment. This is  numerically demonstrated by displaying the  mean absolute error (MAE) between the target eigenstate and the DESD steady state for node and hyperedge signals on: (a) the contact network ($T = 5$, $\sigma = 10$) and (b) the brain network ($T = 10$, $\sigma = 20$). Results are averaged over $30$ head-tail random assignments.
    }}
	\label{MAERobust}
\end{figure*}

\subsection{Empirical validation on contact and biological hypergraphs}
In order to demonstrate the relevance of our findings for real hypergraphs, we have considered two higher-order datasets: a high-school face-to-face contact hypergraph (\textit{contact-high-school})~\cite{Mastrandrea-2015-contact}, and a functional brain hypergraph constructed from resting-state fMRI data (\textit{ABIDE}) using the Schaefer 200-region, 7-network atlas~\cite{di2014autism,schaefer2018local,abraham2014machine}. In the contact data, each interaction is encoded as a directed hyperedge by randomly assigning participants to tail ($T_h$) and head ($H_h$) sets.  {This assignment does not reflect an intrinsic directionality in the data, but serves as a neutral embedding that allows the DESD dynamics to be defined.} For the brain data, hyperedges are constructed by thresholding functional connectivity and grouping regions with similar connection strengths, preserving modular organization while encoding multi-region interactions. In both cases, we focus on connected sub-hypergraphs.  

For the contact network, we consider a sub-hypergraph formed by four course-related groups $\{\mathrm{PC}, \mathrm{PC}^*, \mathrm{MP}^*1, \mathrm{MP}^*2\}$. The Dirac spectrum exhibits several isolated eigenvalues. Thus,  targeting different eigenstates  yields  distinct topological synchronization states  (Fig.~\ref{Fig.6}), where the four course groups partition differently, demonstrating the ability of DESD to design different dynamical partitions of the hypergraph.

For the brain hypergraph, sub-hypergraphs from three resting-state systems—Default Mode, Visual, and Somatomotor—are analyzed. Also in this case, targeting isolated eigenvalues allows us to design different topological cluster synchronization states that display different correlations of the dynamics  with the known functional organization  clusters (see Fig.~\ref{Fig.7}).

While different choices of these sets will lead to different eigenstates of the Dirac Hamiltonian, and thus different topological cluster states, here we show that the convergence of the DESD to such topological cluster states does not depend on the choice of the head-tail sets. To this end, we repeated the integration of the DESD dynamical equation of motion over  30 realizations for each empirical system with random head-tail assignments. The mean absolute error (MAE) between the predicted and simulated cluster states remains small across realizations, as shown in Fig.~\ref{MAERobust}.}

\section{Discussion}
\leavevmode\indent

In this work, we have introduced a general framework for higher-order topological dynamics on directed hypergraphs, addressing a key limitation of existing approaches that are restricted to undirected networks and simplicial complexes. By  formulating an algebraic topology framework based on the Dirac operator of oriented hypergraphs, we establish a unified description of coupled dynamics on nodes and hyperedges, enabling the emergence of topological Dirac cluster synchronization states. These states dynamically partition the hypergraph into different clusters of nodes and different clusters of hyperedges where nodes (or hyperedges) belonging to the same cluster oscillate in unison.

A central conceptual advance of our approach is the demonstration that synchronization patterns can be designed and selected purely through spectral targeting, without modifying the underlying hypergraph structure. Our framework leverages isolated eigenstates of the Dirac Hamiltonian as dynamical attractors, providing a direct and interpretable link between higher-order topology and accessible collective states. Specifically, we show that a single directed hypergraph can support multiple coexisting, spectrally encoded dynamical patterns, each associated with a distinct isolated eigenmode.

Our theoretical predictions are supported by simulations on both synthetic hypergraph block models and empirical datasets, including contact and brain functional networks. These results demonstrate that spectral selection remains effective in realistic, heterogeneous systems, highlighting the robustness and practical relevance of the proposed mechanism.

Overall, our findings establish a  theoretical framework for controlling higher-order dynamics, one that is spectral, topology-driven, and non-invasive, opening the way to applications in complex systems where multi-body and directed interactions play a central role. Future directions include extending the framework to weighted and adaptive hypergraphs, where the interplay between topology and dynamics may further enhance controllability of higher-order topological dynamics.

\section{Methods}
\leavevmode\indent

\subsection{Linear stability analysis}
The linear stability analysis establishes  the stability of the solution $\Psi\simeq \Psi_{\parallel}$ given by Eq.(\ref{eq:synchr_dyn}).
To this end, we extend results obtained in the context of single networks \cite{zaid2025designing,millan2019synchronization,hong2005collective} and we linearize   the dynamical equation for  $\Psi_{\perp}$, given by Eq.(\ref{eq:Psi_dec}), obtaining
\bea
        \dot\Psi_{\perp}=&\Omega_{\perp}-\sigma\, (E-\bar E I)^{2}\Psi_{\perp}
\eea
which has the solution
\bea
        \Psi_{\perp}=&\frac{\Omega_{\perp}}{\sigma(E-\bar E I)^{2}}(1-e^{-\sigma(E-\bar E I)^{2}t})+\Psi_{\perp}(0)e^{-\sigma(E-\bar E I)^{2}t}.
\eea
To predict the linear stability of the  DESD synchronized state, we require that the ``roughness" $W^2$, given by 
\bea \label{eq:W2}
    W^2=\Avg{\|\Psi_{\perp}\|^2},
\eea
 remains bounded  in the limit $t\to \infty$ and  $\mathcal{N}\to \infty$.
In the limit $t\to \infty$, assuming  $\mathcal{N}=N_0+N_1\gg1$, and indicating with $\rho(E)$ the density of states of the Dirac Hamiltonian, we obtain that $W^2$ can be expressed as 
$W^2=W^2_{+}+W^2_{-}$  with 
\bea
W_{+}=\int_{E>\bar{E}+\Delta_{+}}\frac{\rho(E)}{\sigma^2(E-\bar{E})^4},\quad W_{-}=\int_{E<\bar{E}-\Delta_{-}}\frac{\rho(E)}{\sigma^2(E-\bar{E})^4},
\eea
where $\Delta_{+}=\min_{E>\bar{E}}(E-\bar{E})$ and $\Delta_{-}=\min_{E<\bar{E}}(\bar{E}-E)$. Therefore, if $\lim_{\mathcal{N}\to \infty}\Delta_{\pm}>0$, indicating that the eigenstate $\bar{E}$ is isolated, $W^2$ does not diverge. It follows that if $\bar{E}$ is an isolated eigenstate, the DESD synchronized state is linearly stable.  Note that in addition to this analytical linear stability analysis,  we have also numerically observed that if the selected eigenstate $\bar{E}$ is isolated, the corresponding DESD synchronized state has a large basin of attraction. Indeed, provided that $\sigma$ is large enough, the DESD synchronized state is typically achieved starting from arbitrary random initial conditions. By following steps similar to those discussed in the context of simple networks in Ref.\cite{zaid2025designing}, this linear stability analysis also  establishes  the stability of DESD corresponding to the choice of eigenstates ${\bar{E}}$ in the bulk of the spectrum  depending on the spectral properties of the hypergraph.

\subsection{Simulation setup and numerical implementation}
All simulations of the Dirac–Equation Synchronization Dynamics (DESD) were performed using a fourth–order Runge–Kutta integration scheme with a step size of $\Delta t = 10^{-3}$ until convergence. Unless otherwise stated, the coupling constant was set to $\sigma = 20.0$, the mass parameter to $m = 1.0$, and the intrinsic frequencies $\Omega_i$ were generated using \texttt{np.random.rand()} to produce uniformly distributed values in $[0,1]$. Synthetic directed hypergraphs were generated using a Hypergraph Stochastic Block Model (HSBM) with $V = 200$ nodes and four clusters. All analyses were implemented in Python 3.12 using \texttt{NumPy}, \texttt{SciPy}, and \texttt{NetworkX}.

\section*{Data availability}
The source data underlying Figs. 2–9, together with the processed hypergraph datasets used in this study, are available in Zenodo \cite{Guo2026Zenodo} at \url{https://doi.org/10.5281/zenodo.21884946}
And the raw empirical data are publicly accessible from the original sources: the High-School Contact Network (\url{http://www.sociopatterns.org/datasets/})\cite{Mastrandrea-2015-contact} and the ABIDE resting-state fMRI repository (\url{http://fcon_1000.projects.nitrc.org/})\cite{di2014autism}.

\section*{Code availability}
The Python code used in this study has been permanently archived in Zenodo
\cite{Guo2026Zenodo} and is available at
\url{https://doi.org/10.5281/zenodo.21884946}.
All analyses were performed using Python 3.12.

\bibliographystyle{apsrev4-2}
\bibliography{Directed_DESD}

\section*{Founding Statement}
This research utilised Queen Mary's Apocrita HPC facility, supported by QMUL Research-IT. http://doi.org/10.5281/zenodo.438045. This research was supported by  National Natural Science Foundation of China (T2422010 and 62172170) and “the Fundamental Research Funds for Central Universities”.

\section*{Author contributions}
 G.B. and X.L. conceived the study and designed the overall research plan.  
 G.B. and A.A.A.Z. developed the theoretical framework and carried out the analytical derivations.  
Y.G. and A.A.A.Z. performed the numerical simulations and data analysis.  
 G.B. and X.L. supervised the project and contributed to the interpretation of the results.  
All authors discussed the findings, contributed to the manuscript preparation, and approved the final version.

\end{document}